\documentclass[aps,prl,twocolumn,superscriptaddress,showpacs,amsmath,amssymb]{revtex4-1}

\usepackage{graphicx}% Include figure files
\usepackage{dcolumn}% Align table columns on decimal point
\usepackage{bm}% bold math
\usepackage{hyperref}% add hypertext capabilities

\begin{document}

%%%%%%%%%%%%%%%%%%%%% my commands %%%%%%%%%%%%%%%%%%%%%%%%

\newcommand{\fig}[1]{Fig.~\ref{#1}}

\newcommand{\lut}{LuNi$_2$B$_2$C}
\newcommand{\vsi}{V$_3$Si}
\newcommand{\bak}{Ba$_{0.6}$K$_{0.4}$Fe$_2$As$_2$}
\newcommand{\feco}{Ba(Fe$_{0.926}$Co$_{0.074}$)$_2$As$_2$}

\newcommand{\mfc}{$M_{FC}$}
\newcommand{\mzfc}{$M_{ZFC}$}

\newcommand{\tc}{$T_c$}
\newcommand{\hc}{$H_{c1}$}
\newcommand{\hcc}{$H_{c2}$}

%%%%%%%%%%%%%%%%%%%%%%% end of my commands %%%%%%%%%%%%%%%%%%

\title{Anomalous Meissner effect in pnictide superconductors}

\author{R.~Prozorov}
\email[Corresponding author: ]{prozorov@ameslab.gov}
\affiliation{Ames Laboratory and Department of Physics \& Astronomy, Iowa State University, Ames, IA 50011}

\author{M.~A.~Tanatar}
\affiliation{Ames Laboratory and Department of Physics \& Astronomy, Iowa State University, Ames, IA 50011}

\author{Bing~Shen}
\affiliation{Institute of Physics, Chinese Academy of Sciences, Beijing 100190, China}

\author{Peng~Cheng}
\affiliation{Institute of Physics, Chinese Academy of Sciences, Beijing 100190, China}

\author{Hai-Hu~Wen}
\affiliation{Institute of Physics, Chinese Academy of Sciences, Beijing 100190, China}

\author{S.~L. Bud'ko}
\affiliation{Ames Laboratory and Department of Physics \& Astronomy, Iowa State University, Ames, IA 50011}

\author{P.~C.~Canfield}
\affiliation{Ames Laboratory and Department of Physics \& Astronomy, Iowa State University, Ames, IA 50011}

\date{\today}

\begin{abstract}

The Meissner effect has been studied in \feco~ and \bak~ single crystals and compared to well known, type-II superconductors \lut~ and \vsi. Whereas flux penetration is mostly determined by the bulk pinning (and, perhaps, surface barrier) resulting in a large negative magnetization, the flux expulsion upon cooling in a magnetic field is very small, which could also be due to pinning and/or surface barrier effects. However, in stark contrast with the expected behavior, the amount of the expelled flux increases almost linearly with the applied magnetic field, at least up to our maximum field of 5.5 T, which far exceeds the upper limit for the surface barrier. One interpretation of the observed behavior is that there is a field-driven suppression of magnetic pair-breaking.
\end{abstract}


% 74.25.Ha Magnetic properties including vortex structures and related phenomena
% (for vortices, magnetic bubbles, and magnetic domain structure, see 75.70.Kw)

% 74.25.Op Mixed states, critical fields, and surface sheaths

% 74.25.N- Response to electromagnetic fields

% 74.70.Xa Pnictides and chalcogenides

% 74.70.Ad Metals; alloys and binary compounds (including A15, MgB2, etc.)

\maketitle

In textbooks \cite{Gennes1966,Tinkham1996}, the Meissner effect \cite{Meissner1933} is considered to be the definitive mark of bulk superconductivity. In practice, however, there are various factors that determine the behavior of a real, finite specimen in a magnetic field \cite{Gennes1966,Campbell1972,Clem1974,Brandt1995}. Nevertheless, in all cases reported so far, the following characteristic behavior has been observed: magnetization measured as a function of an applied magnetic field after cooling the sample in \emph{zero} field is negative and linear in field up to a characteristic penetration field, $H_p$. Above this field, Abrikosov vortices penetrate the sample and magnetization amplitude decreases \cite{Clem1974}. The value of $H_p$ depends on various parameters, such as sample shape, surface quality, anisotropy and even pairing symmetry, and ranges between the first critical field, $H_{c1}$, and the thermodynamic critical field, $H_c$. For example, in the simplest case without demagnetization when a magnetic field is parallel to the sample surface with characteristic roughness $\sigma$, the Bean-Livingston barrier \cite{Bean1964,Gennes1966,Clem1974} field is given by \cite{Genenko2005},

\begin{equation}\label{eq1}
H_p = \frac{\phi_0}{4\pi\lambda\sigma}\ln{\frac{e\sigma}{\xi}}
\end{equation}

\noindent which ranges between $H_{c1} \leq H_p \leq H_c$ for $\lambda \leq \sigma \leq \xi$. In the field - cooled experiment, the resulting magnetic moment at low temperatures depends on the competition between Meissner expulsion, temperature-dependent pinning strength and surface barrier effects. (Note that we use ``field - cooled'' (FC) to indicate the process when the measurements are taken upon cooling, sometimes denoted as FCC. This may somewhat differ from cooling in an applied magnetic field and measuring upon warming, FCW \cite{Clem1993}). In all cases, however, this resulting moment would first become increasingly negative in fields comparable to the characteristic field discussed earlier and then will start to increase reaching zero at the second critical field, $H_{c2}$. It has been shown that the surface barrier may play an important role in determining the irreversibility in unconventional superconductors close to \tc~ \cite{Konczykowski1991,Burlachkov1993} and may itself be enhanced due to Andreev bound states \cite{Iniotakis2008,Leibovitch2009}. When the applied magnetic field is decreased from a large value, $H \gg H_p$ or the sample is cooled at any field, the barrier actually compensates for the Meissner expulsion \cite{Clem1974,Burlachkov1993}. In real samples with pinning, the competition of temperature-dependent critical current, surface barrier and \hc~ determines the value of the field-cooled magnetization, \mfc, that peaks approximately at $H_p$ and its value is greatly reduced compared to the theoretical Meissner expulsion, $4\pi M_{FC,ideal} =  - HV$, where $V$ is the sample volume. Sadly, this renders standard, field-cooled Meissner effect measurements of little value for the estimation of the ``superconducting fraction''. On the other hand, studies of the field-cooled magnetization of exotic superconductors can reveal new phenomena when compared to the conventional materials.

Here we report an unusual Meissner effect in which the negative field-cooled magnetization continues to become increasingly negative with increasing applied field up to our maximum fields (of 5.5 T), way past the estimated first or thermodynamic critical fields. One possible interpretation of these data, suppression of the magnetic pair-breaking, is discussed, but completely different mechanisms related to the peculiarities of iron - based superconductors might be involved.

\begin{figure}[tb]
\includegraphics[width=8.5cm]{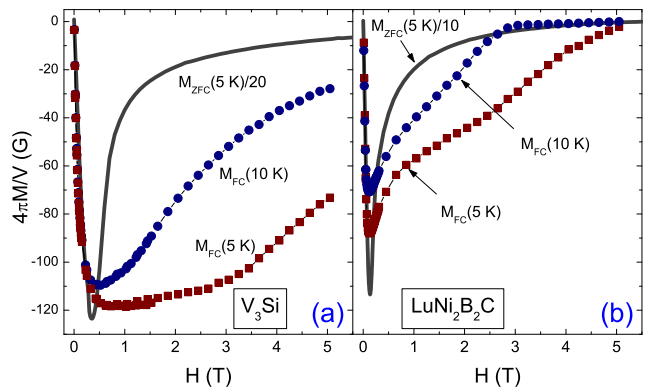}
\caption{Magnetic moment measured after field-cooling, \mfc~ (symbols), and zero-field-cooling, \mzfc~ (solid lines) protocols in single crystals of (a) \vsi~ and (b) \lut. For \mfc$(H)$, each data point was obtained in a separate FC experiment in a constant magnetic field. ZFC curves were scaled by a factor of 20 and 10, respectively.}
\label{fig1}
\end{figure}
%

% experimental

The experiments were performed on single crystals of electron and hole - doped BaFe$_2$As$_2$. Optimally doped single crystals of \feco~(FeCo122) \cite{Ni2008} and \bak~(BaK122) \cite{Luo2008} were grown out of FeAs flux using high temperature solution growth techniques. For comparison, high quality low-pining single crystals of known non-magnetic type-II superconductors, \vsi~ \cite{Samsonov1956} and \lut~ \cite{Canfield1998}, were measured on the same \emph{Quantum Design} MPMS unit following the same protocols. All samples were slab-shaped with dimensions of the order of 2 mm in the $ab-$plane and 0.1-0.3 mm thickness. Several samples of each kind and of different geometries were measured and here we report results obtained on representative crystals. In this work we focus on the case when the magnetic field was oriented along the $ab-$plane to minimize demagnetization effects, but similar results were obtained for the magnetic field along the short dimension, which is along the crystallographic c-axis. Also, similar results were obtained by using a Vibrating Sample Magnetometer (VSM) in applied magnetic fields up to 9 T. To facilitate the comparison, magnetization is presented in gauss by calculating $4 \pi M/V$, where $M$ [emu] is the measured magnetic moment and $V$ [cm$^3$] is the sample volume.

% results

Figure \ref{fig1} shows the data for two well-studied, conventional (as far as vortex behavior is concerned) type-II superconductors with $T_c$ of the same order as the pnictide systems studied here. In both panels the symbols show a magnetic moment, $M_{FC}$, measured after cooling in a constant magnetic field to a fixed temperatures of 10 and 5 K. \emph{Each} data point is the result of a \emph{separate} field-cooling procedure. For comparison, standard magnetization curves obtained after cooling in zero-field to 5 K are shown by solid lines. Note the scaling by a factor of 20 and 10 in panels (a) and (b), respectively. As discussed in the introduction, such behavior is expected for a regular type-II superconductor. Namely, above $H_p$, both ZFC and FC magnetic moments decrease in magnitude indicating an increasing density of vortices. Note that despite considerable differences in \mzfc~ and upper critical fields \cite{Metlushko1997,Khlopkin1999}, the maximum amplitude of the negative field-cooled magnetization of these two systems is quite similar,  but even they have far from the ideal value of the Meissner expulsion.

\begin{figure}[tb]
\includegraphics[width=8.5cm]{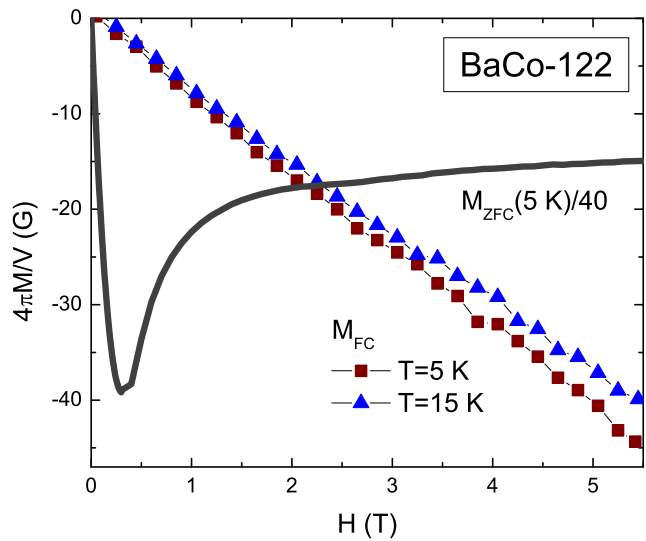}
\caption{Comparison of \mfc~ (symbols) (sampled at 5 K and 15 K) and \mzfc/40~ (line) measured by sweeping a magnetic field at 5 K in a FeCo-122 crystal.}
\label{fig2}
\end{figure}

Figure \ref{fig2} shows similar data obtained on a FeCo-122 crystal. Measurements of \mfc, shown at 5 and 15 K, reveal a striking difference in comparison with Fig.~\ref{fig1}. The magnetic moment increases in amplitude becoming more negative almost linearly in field all the way up to our largest applied field of 5.5 T. The \mzfc$(H)$ curve, on the other hand, exhibits standard behavior, similar to the curves shown in Fig.~\ref{fig1}.

\begin{figure}[tb]
\includegraphics[width=8.5cm]{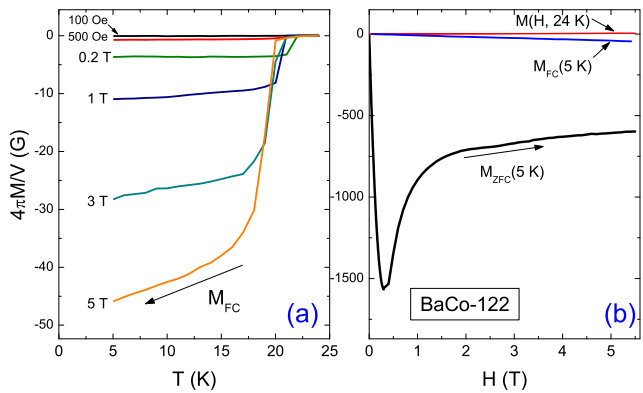}
\caption{(a) magnetization measured upon field-cooling at different values of the applied magnetic fields in a FeCo-122 crystal. (b) measured \mzfc~ and \mfc~ at 5 K compared to the magnetization above $T_c$ where \mfc=\mzfc.}
\label{fig3}
\end{figure}

To further illustrate this unusual effect, \fig{fig3}(a) shows several \mfc$(T)$ curves measured at different applied fields in a FeCo-122 single crystal.
The magnetic moment is clearly much more negative at higher fields, except for in the vicinity of \tc, as expected. To examine whether this effect is related to the normal-state response, \fig{fig3}(b) shows measurements of \mfc~ at 5 K and 24 K, i. e., just above \tc. An  unscaled \mzfc~ curve at 5 K is also shown to illustrate the relative magnitudes. As expected and known for the Fe-based superconductors, $M(H)$ above \tc~ is weakly paramagnetic and, therefore, the observed diamagnetism must come from the superconducting state itself.

\begin{figure}[tb]
\includegraphics[width=8.5cm]{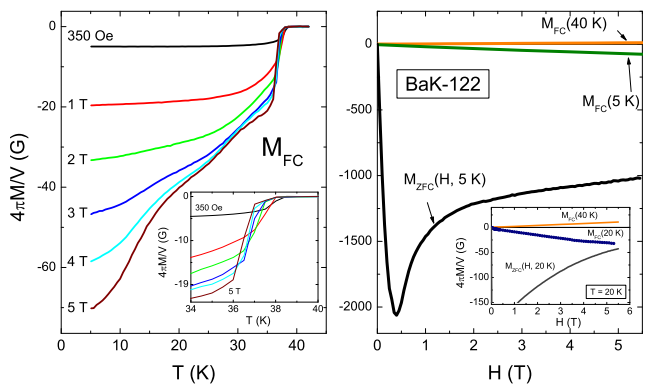}
\caption{(a) field-cooled magnetization vs. temperature measured at different magnetic fields in a BaK-122 crystal. (b) measured \mzfc~ and \mfc~ at 5 K compared to the magnetization above \tc. Inset: similar data for $T=20$ K.}
\label{fig4}
\end{figure}

Similar behavior is observed for BaK-122 crystals. Figure~\ref{fig4}(a) shows results of field-cooling experiments as a function of temperature where each curve was obtained by measuring in a constant field indicated in the figure. Figure~\ref{fig4}(b) shows the magnetic field dependence of \mfc~ at 5 K along with the $M(H)$ curve measured at 40 K, which is just above \tc~ as well as unscaled \mzfc~ at 5 K, for comparison. The inset shows similar data for 20 K indicating that the relative strength of the effect, i.e., magnitudes of \mfc~ vs. \mzfc, become even more pronounced.
Note that above \tc, \mzfc~ and \mfc~ experiments produce the same reversible curve for all samples. Overall, the observed behavior is very similar to that of FeCo-122 and is at odds with that of conventional type-II superconductors shown in \fig{fig1}.


\begin{figure}[tb]
\includegraphics[width=8.5cm]{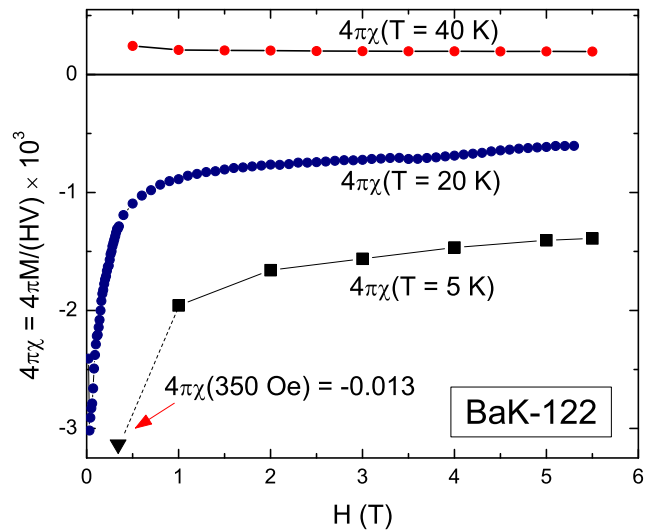}
\caption{Magnetic susceptibility, $4\pi\chi=M/(VH)$, above (40 K) and below (5 K and 20 K) $T_c$ in a BaK-122 crystal. The point at $T = 5$ K and $H=350$ Oe, shown by a triangle, is outside the vertical range with $4 \pi \chi = -0.013$, still much lower than ideal -1.}
\label{fig5}
\end{figure}

To look at the data from a different angle, Fig.~\ref{fig5} shows the magnetic susceptibility $4 \pi \chi = 4 \pi M/(VH)$, which should be equal to -1 in the case of an ideal demagnetization-free superconductor and rapidly decrease in amplitude above the characteristic field $H_p$. Clearly, the field-cooled magnetic susceptibility (shown for 5 K and 20 K) is very small, of the order of $1\times10^{-3}$. It is almost field-independent and negative up to our maximum field. Measurements above \tc~ (at 40 K) reveal field-independent, but positive valued susceptibility, as is expected for a paramagnetic response.

% discussion
The data show a clear difference between conventional type-II superconductors and pnictides. Whereas the values of \tc~ are comparable in all studied samples, the upper critical fields, \hcc, are quite different. In particular, for $H \perp c-$axis, $H_{c2} \sim 55$ T in FeCo-122 \cite{Ni2008,Kano2009}, 80 T in BaK-122 \cite{Altarawneh2008} as opposite to 20 T in \vsi~ \cite{Khlopkin1999} and 9 T in \lut~ \cite{Metlushko1997}. Still, the effect is observed in fields much greater than \hc~, but much lower than \hcc, indicating that this is a property of a robust superconducting state with a fully developed order parameter. The negative magnetization upon field-cooling is determined by the Meissner effect which  has a magnitude that is proportional to the lower critical field, \hc, which is small in pnictides, $H_{c1} \sim 100$ Oe \cite{Gordon2010a}. If the observed effect is due to the surface barrier, then following Eq.~\ref{eq1}, the maximum critical field (for ideally smooth surface, $\sigma \leq \xi$) where it is effective is of the order of the thermodynamic critical field $H_c=\sqrt{H_{c1}H_{c2}} \sim 1$~T if we take the overestimated $H_{c2} \sim 100$ T. In real samples, the surface barrier field would be less effective due to finite roughness and scattering. Therefore, the conventional Bean-Livingston mechanism does not explain our results.

Another scenario is the competition between bulk pinning and the Meissner effect. Upon cooling, the temperature - dependent pinning prevents some vortices from exiting and a characteristic dome-like shape of the magnetic induction is formed with the Meissner expulsion region confined to the sample edges \cite{Dorosinskii1993}. In principle, strong suppression of pinning by a magnetic field could lead to an increase of the negative magnetization. However, this is not observed in most superconductors and direct measurements show that in the discussed temperature and field range, the pinning in pnictides is quite strong and is not particularly different from known materials.

One interesting possibility is that the magnetic field aligns local magnetic moments of iron ions as well as of vacancies and other lattice disturbances introduced by the doping. These moments act as efficient pair-breakers and, therefore, such Zeeman alignment will decrease the pairbreaking that involves flipping the spin of the scattering center. Experiments and theory have suggested that such pairbreaking scattering is very important in iron-based superconductors \cite{Gordon2010,Kogan2009a} and perhaps our results provide further support for this idea. On the other hand, many theoretical and experimental reports indicate importance of magnetic fluctuations in the mechanism of superconductivity in iron-based superconductors. Therefore, it seems that the pairing and the pair-breaking both originate from the same magnetic subsystem and, therefore strength of superconductivity (e.g., transition temperature) should be calculated considering both effects.

The reversible magnetization of a type-II superconductor for $H_{c1} \ll H \ll H_{c2}$ is given by,

\begin{equation}\label{eq2}
M=-\frac{\phi_0}{32\pi^2\lambda^2}\ln{\frac{\eta H_{c2}}{H}}
\end{equation}

\noindent where $\eta \sim 1$ \cite{Kogan2006}. In the strong pair-breaking regime, $\lambda^{-2} \sim \tau_m$, where $\tau_m$ is the magnetic scattering time. Taking into account the weak paramagnetic response and a relatively small magnetic moment per iron, $m \sim 0.87 \mu_B$ \cite{Huang2008a} in BaFe$_2$As$_2$, we can assume a simple correction of the scattering time, $\tau_m = \tau_m(H=0)\exp{(mH/k_BT)}$. With $m/k_B \leq 0.58$ K/T we can therefore write even at our highest fields and lowest temperatures, $\tau_m \approx \tau_m(H=0)(1+mH/k_BT)$ \cite{Kogan2009a}. Therefore, we would expect that the reversible magnetic moment will increase in amplitude as, $-M \sim H/T$. Surprisingly enough, this simple model describes our observations quite well. An almost linear field dependence is evident from the figures \ref{fig2}-\ref{fig4} and a pronounced temperature is observed, especially at higher fields, see Figs.~\ref{fig3}(a) and \ref{fig4}(a). We note that very high upper critical fields of pnictide superconductors are essential for this mechanism to work. Otherwise, the superconducting order parameter gets suppressed and reversible magnetization decreases rapidly, see Eq.~\ref{eq2}.

In conclusion, we report on the anomalous field-cooled magnetization in BaFe$_2$As$_2$ - based superconductors. The magnetic moment becomes progressively more negative as a function of an applied magnetic field that  significantly exceeds the thermodynamic critical field. It is proposed that the observed behavior can be explained by a field-induced reduction of magnetic pairbreaking. In addition to the presence of natural magnetic scatterers in iron - pnictides, large values of \hcc~ make it possible to observe the effect.

%Acknowledgement
We thank V.~G.~Kogan, J.~R.~Clem, A.~Gurevich, and L.~Burlachkov for useful discussions and D.~K.~Christen for providing \vsi~ crystal. The work at Ames Laboratory was supported by the U.S. Department of Energy, Office of Basic Energy Sciences, Division of Materials Sciences and Engineering under contract No. DE-AC02-07CH11358. The work in China was partially supported by the Natural Science Foundation of China, the Ministry of Science and Technology of China (973 Projects No.2011CB605900). R.P. acknowledges support from the Alfred P. Sloan Foundation.

\end{document}